\documentclass[nonacm,sigplan]{acmart}

\usepackage[]{hyperref}

\usepackage{threeparttable}
\usepackage{multirow}
\usepackage{listings}

\colorlet{mygray}{black!30}
\colorlet{mygreen}{green!60!blue}
\colorlet{mymauve}{red!60!blue}
\def\codename{HiCCL}

\begin{document}

\title{HiCCL: A Hierarchical Collective Communication Library}

\author{Mert Hidayetoglu}
\affiliation{%
  \institution{Stanford University}
  \country{CA, USA}
}
\author{Simon Garcia de Gonzalo}
\affiliation{%
  \institution{Sandia National Laboratories}
  \country{NM, USA}
}
\author{Elliott Slaughter}
\affiliation{%
  \institution{SLAC National Accelerator Laboratory}
  \country{CA, USA}
}

\author{Pinku Surana}
\affiliation{%
  \institution{SLAC National Accelerator Laboratory}
  \country{CA, USA}
}

%\author{Patrick McCormick}
%\affiliation{%
%  \institution{Los Alamos National Laboratory}
%  \country{NM, USA}
%}

\author{Wen-mei Hwu}
\affiliation{%
  \institution{Nvidia / University of Illinois at Urbana-Champaign}
  \country{IL, USA}
}
\author{William Gropp}
\affiliation{%
  \institution{University of Illinois at Urbana-Champaign}
  \country{IL, USA}
}
\author{Alex Aiken}
\affiliation{%
  \institution{Stanford University}
  \country{CA, USA}
}
% removed for anonymity

\begin{abstract}
HiCCL (Hierarchical Collective Communication Library) addresses the growing complexity and diversity in high-performance network architectures. As GPU systems have envolved into networks of GPUs with different multilevel communication hierarchies, optimizing each collective function for a specific system has become a challenging task. Consequently, many collective libraries struggle to adapt to different hardware and software, especially across systems from different vendors. HiCCL's library design decouples the collective communication logic from network-specific optimizations through a compositional API. The communication logic is composed using multicast, reduction, and fence primitives, which are then factorized for a specified network hieararchy using only point-to-point operations within a level. Finally, striping and pipelining optimizations applied as specified for streamlining the execution.
 
 Performance evaluation of HiCCL across four different machines—two with Nvidia GPUs, one with AMD GPUs, and one with Intel GPUs—demonstrates an average 17$\times$ higher throughput than the collectives of highly specialized GPU-aware MPI implementations, and competitive throughput with those of vendor-specific libraries (NCCL, RCCL, and OneCCL), while providing portability across all four machines.
\end{abstract}

\maketitle % should come after the abstract
\pagestyle{plain} % should come right after \maketitle

\section{Introduction}

%Collective communications involve all GPUs in a system in various ways .
Communication libraries in high-performance computing (HPC) for scientific applications and machine learning offer \textit{collectives}---standard functions such as Scatter, Reduce, and All-Reduce \cite{MPICH,OpenMPI,thakur2005optimization} that involve coordinated communication across processors. Our evaluation on leadership-class GPU systems revealed that the throughput of available library collectives is either not optimized in GPU-aware MPI implementations or have limited cross-vendor portability in NCCL, RCCL, and OneCCL. These performance and portability issues  motivated us to develop a library design capable of producing throughput-optimized collectives portable across diverse network architectures and GPU programming models. However, there are many collective functions, each requiring a unique optimization strategy depending on the network architecture~\cite{wilkins2023generalized}, making it unproductive to hand optimize a collective for one system and then repeat the process when porting to another system. %In short, optimizing collective communications across a diverse set of modern systems is challenging.

Modern systems typically feature a network comprised of multiple compute nodes each housing multiple GPUs~\cite{Frontier, Sunspot, Perlmutter, Delta}, which results in a nonuniform interconnect architecture among the GPU endpoints. Specifically, the network is hierarchical: GPUs within the same node communicate via a high-throughput intranode network, while GPUs on different nodes communicate through a lower-throughput network that links the nodes. Consequently, the bandwidth available for GPU communication is influenced by their placement within the hierarchical network architecture~\cite{siefert2023latency}.
%This hierarchy generally has at least three levels: NUMA domains when there are multiple GPUs on a die, within a node, and across nodes is currently typical.
This hierarchy can extend to multiple levels, such as GPU dies in a device, devices across NUMA nodes, or nodes within racks, where communication bandwidth decreases at higher levels of the hierarchy, making data transfer between more “distant” GPUs costlier.
Communication bandwidth decreases at higher levels of the hierarchy, making data transfer between more ``distant'' GPUs costlier. Efficient hierarchical communication relies on having lower volume at higher levels of the hierarchy~\cite{hidayetouglu2019memxct, bienz2021modeling, lockhart2023characterizing}, which is 
currently achieved with carefully hand-designed and optimized implementations of collectives.
%achieved by factoring a collective communication operations into multiple components, each tailored to a specific level of the network hierarchy, and pipelining these stages for overlapping the communications.
%The structure of the network hierarchy across GPUs depends mostly on the node architecture, where GPUs communicates with a higher bandwidth within nodes than that of across nodes.
%Hierarchical communication is about minimizing the communication volume on higher levels by rerouting (and reusing) the messages in lower levels in an optimized way.
%across the levels in the GPU network hierarchy~[SC20].

%Recent work applies hierarchical communications for optimizing collective communications. They they either work on a single node [MSCCL] node or CPU-only [Xi] settings. But most importantly, they rely on a single library, limiting performance portability.
Our aim is to provide both high throughput
and performance portability for collective communication operations across diverse hierarchical networks with GPUs of different vendors\footnote{Today's major GPU vendors are Nvidia, AMD, and Intel.} while also automating much of the process of building optimized collectives. We present a method for constructing collective communication operations that achieves these goals in two steps:
First, we specify the communication pattern for a  collective operation as a composition of multicast, reduction, and fence primitives (Section~\ref{sec:design}).
Second, we identify and implement a number of optimizations that can be mechanically applied to the given composition (Section~\ref{sec:optimization}).  
These optimizations include striping communication across multiple network interface cards (NICs), pipelining communication in multiple stages, and whether to use a customized tree, ring, or hybrid virtual topology to best match the target network hierarchy.
Combining these optimizations with a machine-specific description that fills in important constants (e.g., the number of levels in the hierarchy and the base communication library to use for each level), users can create highly optimized collective operations tuned to a specific network.
%decoupling the collective communication design from its machine-specific optimization and implementation. To this extent, we propose a compositional API that allows the user to build desired collective function using two fundamental data movement schemes and a fence primitive. %three primitive functions 
%(Section~\ref{sec:primitives}). The second goal is describing the hierarchy so that optimizations can be automatically applied to any collective function that is composed with the proposed API. We achieved this with a simplified machine model (Section~\ref{sec:machine}) that abstracts most of the architectural detail but captures the necessary information to apply hierarchical optimizations. 
When porting between machines, only the machine description
needs to change; the specification of the logic of the collective operation can be automatically optimized for the target network using the new machine description.

%but the machine description for optimal communication synthesis. 

We implement our optimized collectives relying only on point-to-point communication functions of standard communication libraries and, when
available, vendor-provided capabilities for a specific machine.
This implementation strategy gives us both maximum flexibility to use the best point-to-point primitives for a specific
level of a network hierarchy while also ensuring maximum
portability (Section~\ref{sec:impl}).

We have implemented these ideas in a hierarchical collective communication library, \codename{}, which provides an API to build collective functions and apply hierarchical optimizations. Moreover \codename{} provides performance portability across systems of different shapes and sizes and of different vendors. We make the following contributions:
\begin{itemize}
    \item We introduce a machine-agnostic specification of collective functions using multicast, reduction, and fence primitives. We show these primitives are sufficient to express all collective functions in the MPI standard and their alternative implementations. %With these primitives, the user can compose complicated patterns without worrying about their optimization for a specific system.
    \item We identify a set of unified hierarchical optimizations that are applicable to any collective function composed with the proposed primitives.
    We show how the optimizations adapt to different, modern GPU systems, and are sufficient to saturate the throughput of the various networks.
    %These optimizations can adapt different modern GPU systems and sufficient enough to saturate the collective communication throughput on different network architectures.
    \item We introduce HiCCL\footnote{https://github.com/merthidayetoglu/HiCCL}: a hierarchical communication library which integrates multiple communication capabilities without relying on existing collective functions. We demonstrate HiCCL's performance portability by matching or outperforming available MPI and vendor-provided libraries on different systems with Nvidia, AMD, and Intel GPUs.
\end{itemize}
%We ensure that our library can easily adapt to the current and future exascale systems with additional hierarchies within and across nodes.

We evaluate \codename{} on four current HPC systems---Delta~\cite{Delta} and Perlmutter~\cite{Perlmutter} with Nvidia GPUs, Frontier~\cite{Frontier} with AMD GPUs, and Aurora~\cite{Sunspot} with Intel GPUs---and on eight collective functions that are listed in Table~\ref{tab:collectives}. HiCCL achieves 17$\times$ geomean speedup over the native MPI implementations across all platforms. Furthermore, it delivers competitive (1.27$\times$ against NCCL/RCCL), and in some cases superior (12.1$\times$ against OneCCL), throughput compared to the libraries provided by the GPU vendors.% \codename{} demonstrates substantial speedup over native MPI implementations on each system and competitive throughput against vendor-provided libraries, if not more. %Porting across different system architectures does not require changing more than a few lines of HiCCL code. % As a result, \codename{} outperforms all other already available alternatives (MPI, NCCL, RCCL, OneCCL) on our test systems.

\section{Background}

%Figure~\ref{fig:overview_bandwidth} shows the peak bandwidth performance of the proposed library on a few selected collective communication functions. In this work, we test the performance on three contemporary systems with different different architectures (See Table~\ref{tab:systems}). The figure shows the breakdown of performance improvements with the hierarchical communication and pipelining optimizations.

%To assess the performance we take three baselines: 1) the corresponding 1-1 performance of the MPI and NCCL collective function calls, 2) theoretical upper bound based on the NIC bandwidth per each node, and 3) corresponding one-factor (direct) implementation with MPI or NCCL (we choose the faster one). Our results show an obvious performance win compared to the MPI and NCCL implementations. We will discuss the details in Section~\ref{sec:evaluation}.

This section provides an overview of collective functions and hierarchical communication on multi-GPU, multi-NIC node architetures.

% \usepackage{multirow}
% \usepackage{graphicx}
\begin{table}[t]
\centering
\caption{Collectives of MPI, NCCL, RCCL and OneCCL.}
\label{tab:collectives}

\resizebox{\columnwidth}{!}{%
\begin{tabular}{llll}
\textbf{Collective} & \textbf{MPI} & \textbf{NCCL / RCCL} & \textbf{OneCCL} \\ \hline
Scatter             & \texttt{MPI\_Scatter}       &                        &  \\
Broadcast           & \texttt{MPI\_Bcast}         & \texttt{ncclBcast}     & \texttt{ccl::broadcast}  \\ 
Gather              & \texttt{MPI\_Gather}        &                        & \\
Reduce              & \texttt{MPI\_Reduce}        & \texttt{ncclReduce}    & \texttt{ccl::reduce} \\ 
All-to-all          & \texttt{MPI\_Alltoall}      &                        & \texttt{ccl::alltoall} \\
All-gather          & \texttt{MPI\_Allgather}     & \texttt{ncclAllgather} & \texttt{ccl::allgatherv} \\
Reduce-scatter      & \texttt{MPI\_Reduce\_sc.}   & \texttt{ncclReduceSc.} & \texttt{ccl::reduce\_sc.} \\
All-reduce          & \texttt{MPI\_Allreduce}     & \texttt{ncclAllreduce} & \texttt{ccl::allreduce}  \\ 
\end{tabular}%
}
%{\raggedright  \footnotesize NCCL and RCCL has the same API except they use CUDA and HIP streams, respectively.\par}
\end{table}

\subsection{Conventional Libraries and Collective Functions}

MPI~\cite{snir1998mpi} is a well-established standard for message passing in distributed computing, with numerous implementations~\cite{MPICH,OpenMPI,SpectrumMPI, panda2021mvapich}. GPU-aware implementations of MPI are available for almost all systems, including our test systems.  NCCL~\cite{jeaugey2017nccl} is a vendor-provided library, specifically developed for Nvidia GPUs and based on CUDA. RCCL~\cite{amd_rccl} is an analogous library provided by AMD, mirroring NCCL's API for compatibility but based on HIP. OneCCL~\cite{oneccl} is Intel's collective communication library.
The NCCL, RCCL and OneCCL libraries offer the standard collective communication functions in Table~\ref{tab:collectives}, while MPI's API offers additional functions not show~\cite{gropp1999using}. %On the other hand, NCCL, RCCL, and OneCCL are limited to those in Table~\ref{tab:collectives}. %Conversely, \codename{} has a composable interface which optimizes all collective patterns presented in Table~\ref{tab:collectives} and enhances performance on a given GPU network architecture.

%Within nodes, libraries use inter-process communications (IPC). Across nodes, they use specialized NIC hardware and software.

\begin{figure}[t]
\includegraphics[width=\columnwidth]{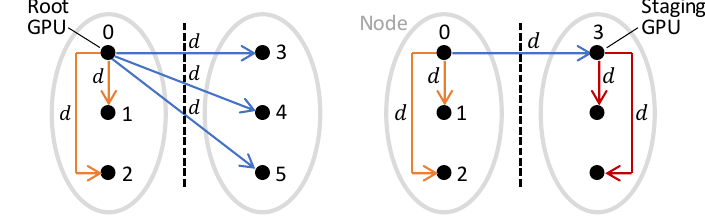}
\centering \\
\small \hspace{0.2cm} (a) Direct \hspace{2.5cm}(b) Hierarchical 
\caption{Broadcasting $d$ bytes across six GPUs with (a) direct and (b) hierarchical ways. Each black dot corresponds to a GPU endpoint. Each set of three GPUs corresponds to a compute node. (a) Direct implementation redundantly moves three copies (blue) of data across nodes. (b) Hierarchical optimization moves a single copy across (blue) nodes, and distribute additional copies within (maroon) nodes.}
\label{fig:hierarchical_broadcast}
\Description[<short description>]{<long description>}
\end{figure}

\subsection{Hierarchical Communication}\label{sec:hierarchical_background}

Hierarchical GPU networks involve groups of GPUs organized into nested, multi-level structures, where groups within the same level share the same number of GPUs. Normally inter-group communication bandwidth is lower than intra-group bandwidth on each level, so the aim of hierarchical communication strategies is to minimize data transfer volumes between groups at all levels. 

%Hierarchical communications assume that the bandwidth between different groups is lower than the bandwidth within the same group, and optimizes collective communications such that the communication volume across groups are minimized.
%To explain, we consider a two-level hierarchy with two multi-GPU nodes with three GPUs each, and a Broadcast of $d$ bytes from the root GPU (rank 0).
As an example, we illustrate a simple two-level hierarchy with two groups of nodes each  with three GPUs.  We describe a process for broadcasting $d$ bytes from GPU 0 to all other GPUs.
Figure~\ref{fig:hierarchical_broadcast}(a) shows {\em direct} communication, where orange represents communication within the {\em root} node (the node with GPU 0)  and blue represents communication across nodes.  All five transfers move all $d$ bytes, which is unnecessary.  In contrast, {\em hierarchical} communication breaks the communication across nodes into two stages as shown in Figure~\ref{fig:hierarchical_broadcast}(b): The first stage sends a single copy between the nodes (blue), and the second stage distributes the data within the sending (orange) and receiving (maroon) nodes. While direct communication has only a single stage where all transfers are done in parallel, in general inter-node bandwidth limitations will result in the direct strategy being much slower than the hierarchical strategy.  Furthermore, pipelining can be used to hide the overhead of multiple stages, a technique further discussed in Section~\ref{sec:pipelining}.% These nodes are connected to the external network with $k$ NICs each, and therefore the bisection bandwidth is assumed to be $$

%a broadcast across these GPUs with direct (point-to-point) communications.
%In two-level GPU communication, intra-node bandwidth hinges on the link speed between GPU pairs, while inter-node bandwidth is contingent on the number of engaged NICs. %Despite existing multi-NIC striping algorithms~\cite{coll2003using},  in our experiments most higher-level communication libraries underutilize the multi-NIC node architecture. To address this, \codename{} incorporates a striping algorithm, achieving speedup  in both point-to-point (P2P) and collective communications, explained further in Section~\ref{sec:design}.  
%This hierarchical optimization assumes greater cost (i.e., slower communication) at higher levels of the hierarchy, such as across nodes, compared to lower levels like within nodes. Effectively, minimizing higher-level traffic through hierarchical factorization.

%To assess \codename{}'s performance, we compare achieved throughput to theoretical upper bounds, depicted in Figure~\ref{fig:overview_bandwidth}. These bounds depend on multi-GPU, multi-NIC node parameters: $p$ (number of GPUs), $g$ (GPUs per node), $n$ (NICs per node), and $f$ (NIC unidirectional bandwidth). Table~\ref{tab:upper_bounds} list these bounds, assuming inter-node communications as the bottleneck and thus disregarding intra-node costs. This subsection further elaborates on collective patterns and their respective theoretical upper bounds.

\begin{figure}[t]
\includegraphics[width=0.9\columnwidth]{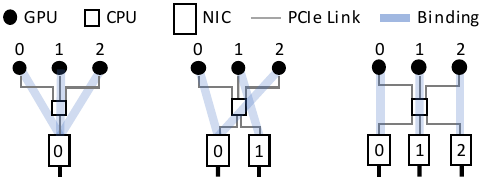}
\centering \\
\small \hspace{-0.1cm} (a) Packed \hspace{1.1cm}(b) Round-robin \hspace{0.95cm} (c) Bijective
\caption{Various associations across $g$ GPUs and $k$ NICs per node ($k\le g$). In our test systems, each GPU is logically binded to a single NIC via (a) packed, (b) round-robin, or (c) bijective associations.}
\label{fig:multi_NIC_node}
\Description[<short description>]{<long description>}
\end{figure}

\subsection{Communications Across Multi-NIC Nodes}\label{sec:multi_NIC}

%In multi-GPU nodes  However, the illustrated scenario in practice would only utilize single NIC, and therefore does not fulfill the full bisection bandwidth across nodes as elaborated next.  

To provide adequate compute-to-communication ratios, multi-GPU nodes now incorporate multiple NICs ~\cite{de2020depth, zimmer2019evaluation}.
%Theoretically, communication across nodes can achieve a total bandwidth of $kf$, with $k$ representing the NIC count and $f$ the NIC bandwidth.
Our experiments with other communication libraries on our test systems show that each process is assigned to a single NIC statically throughout the lifetime of an application. As a result, in the common case where each process manages a single GPU, each GPU is indirectly bound to a single NIC. Consequently, both the direct and hierarchical implementations in Figure~\ref{fig:hierarchical_broadcast} utilize only a single NIC for communication across nodes. We address this limitation by offering multi-NIC striping with HiCCL, detailed in Section~\ref{sec:striping_optimization}. 

When using multiple NICS to handle the communication from a single GPU, 
it is important to understand exactly how that binding is done.   Figure~\ref{fig:multi_NIC_node} shows $g$-to-$k$ bindings within a hypothetical node of $g$ GPUs and $k$ NICs. When $g$ is an exact multiple of $k$, we assign GPUs to NICs in a load-balanced way with (a) packed and (c) bijective mappings, which achieve the full bandwidth across nodes when all GPUs send/receive the same amount of data. However, when $g$ is not a multiple of $k$, we use a (b) round-robin assignment that may lead to load imbalance across the NICs. Consequently, the utilization is only 75\% of the theoretical bandwidth when all GPUs send/receive the same amount of data in Figure~\ref{fig:multi_NIC_node}(b). This under-utilization will have implications in our evaluation in Section~\ref{sec:evaluation_bounds}.

\section{Composition of Collectives}\label{sec:design}

%In this section, we explain how to compose single-step and multi-step collective functions using our proposed fundamental data movement and synchronization primitives.

HiCCL has a compositional design built on three primitives: {\em multicast}, {\em reduction}, and {\em fence}.

\subsection{Collective Primitives}\label{sec:primitives}

%\codename{}'s API allows expressing any collective function in the MPI standard within 3--5 lines of code. \codename{}'s core premise is that by factorization of individual primitives for a described hierarchy \textit{and} streamlining their execution leads to overall optimization of the composed pattern.

%The simplest form of data movement is a point-to-point communication from a single GPU to another, which signifies to a data dependency across the GPUs. Collective communication, however, involves data dependencies among multiple GPUs. If one express a collective functions directly using point-to-point communications, the resulting performance is more likely to be sub-optimal because optimizations on hierarchical networks that are discussed in Section~\ref{sec:hierarchical_background} are missed. Therefore we need to expressing data dependencies across multiple GPUs so that can be exploit for a specific hierarchical network. We propose modified \textit{broadcast} and \textit{reduce} as a fundamental primitives for hierarchical collectives.
The simplest communication building block is point-to-point communication between two GPUs.  Directly expressing collectives using point-to-point semantics often results in sub-optimal performance due to obscuring opportunities for the optimizations discussed in Section~\ref{sec:hierarchical_background}. Therefore, we find it beneficial to 
use three higher-level primitives, which are ultimately implemented using point-to-point operations after optimizations have been applied.

\begin{figure}[b]
\includegraphics[width=\columnwidth]{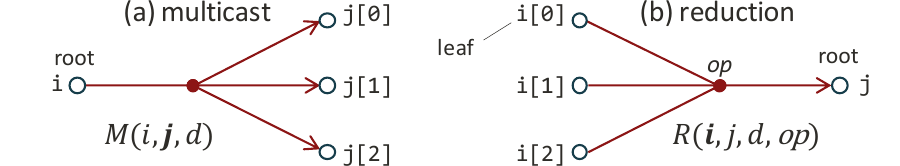}
\centering
\caption{The (a) multicast and (b) reduction primitives form a simple tree structue with one root and multiple leaves.}
\label{fig:primitives}
\Description[<short description>]{<long description>}
\end{figure}

The {\em multicast} primitive  $M(i, \boldsymbol{j}, d)$ expresses a one-to-many dependency between GPUs shown in Figure~\ref{fig:primitives}~(a), where the \textit{root} GPU with {\em rank} $i$ replicates $d$ bytes of data to multiple \textit{leaf} GPUs whose ranks are represented by the vector $\boldsymbol{j}$. The leaf GPUs may be a sparse subset of all GPUs. 

The {\em reduction} primitive $R(\boldsymbol{i}, j, d, \textit{op})$ expresses many-to-one dependencies shown in Figure~\ref{fig:primitives}~(b), where datums of size $d$ at the leaf GPUs $\boldsymbol{i}$ are reduced to a single datum at the root GPU $j$.  The reduction primitive additionally involves a computation \textit{op} such as sum, max, logical or, etc..

%The set of leaf GPUs is a sparse subset of all GPUs (including the root).
If the leaf set involves all GPUs, the multicast and reduction primitives correspond to the traditional Broadcast and Reduce collectives, respectively. On the other hand, when there is a single GPU in the leaf set, the primitives logically simplify to point-to-point communication (with an omitted  unary operation in the case of reduction). 
%In this case, the multicast $M(i, j, d)$ and reduction $R(i, j, d, \textit{op})$ primitives are functionally equivalent because the kernel call (\textit{op}) in the latter is ignored when there is one sender and one receiver. %However, they are broken down into multiple steps in different ways as explained in Section *****.% If the sender and receiver GPUs are the same, the data movement corresponds to a self (in-memory) communication.

\begin{lstlisting} [numbers=none, caption={C++ API for registering primitives.}, label=listing:primitive_API, float=t]

void Comm<T>::add_multicast(T *sendbuf, T *recvbuf, size_t count, int i, std::vector<int> j_vec);
void Comm<T>::add_reduction(T *sendbuf, T *recvbuf, size_t count, std::vector<int> i_vec, int j, HiCCL::op);
void Comm<T>::add_fence();
\end{lstlisting}

 %In the abstract form of $R$ and $B$, the the implementation details are omitted. One more detail to specify the pointers to send or receive buffers at each endpoints of the primitive.
Multiple multicast and reduction reduction primitives can be composed by registering them in a
 a persistent global communicator, \texttt{HiCCL::Comm<T>}; the semantics is that all registered primitives will be executed in parallel. The communicator is templatized over the communicated data type \texttt{T}.
Each primitive registration is made with an invocation of the C++ API shown in Listing~\ref{listing:primitive_API}. The registration functions accept pointers to the communication buffer (\texttt{sendbuf} / \texttt{recvbuf}), number of elements (\texttt{count}), ranks of the sender and receiver GPUs in scalar (\texttt{i} and \texttt{j}), and ranks of the receiver and sender GPUs in vector form (\texttt{j\_vec} and \texttt{i\_vec}) when registering a multicast and reduction primitives, respectively. A reduction operation is expressed by the last argument (\texttt{HiCCL::op}) of the registry function in Listing~\ref{listing:primitive_API}.  %This design allows intuitional programming of collective functions.

\subsection{Single-Step Collectives}\label{sec:collectives}

%As mentioned before, composition of a collective function is done through registering one or multiple primitives into the HiCCL communicator. The ranks of the root and leaf GPUs for each primitive should be expressed according to the desired composition. In designing the collective pattern,
%it should be kept in mind that the primitives will be executed simultaneously. 
 % The designer should consider the collective pattern as the superposition of the primitives that are expressed through the proposed API (Listing~\ref{listing:primitive_API}).
%This design space yield a single-step formulation of collectives, where the registered primitives do not depend on each other.
%All collective functions in the MPI standard except Scan can be expressed in a single step within a loop of 1--3 lines of code.

A single-step collective is composed of one or more primitives that are executed concurrently in a single step. If there are any race conditions between primitives the result is undefined. 
Table~\ref{tab:composition} (Single) shows the realization of the standard collective communication functions using single-step collective functions. For consistency, the largest buffer size is chosen as $dp$, where $p$ is the total number of GPUs. Summations $\Sigma$ express the parallel composition of multiple primitives into a single-step collective (the  vector $\boldsymbol{U}$ represents all participating GPUs). The range of the summations is $(0,p-1)$. The summations (and so the registration of primitives) can be performed in any order. For example, Broadcast and Reduce can be expressed with a single primitive, which directly maps to a single line of code in HiCCL. All-to-all requires $p^2$ point-to-point primitives, and can be expressed with two nested loops and a single call to a primitive in three lines of HiCCL API code. The rest of the single-step collectives require $p$ primitives that can be written using a loop with a single primitive in two lines of API code. 

The top half of Figure~\ref{fig:allreduce}(a) visualizes the composition of a Reduce-scatter on three processes, where the initial data is 3$d$ bytes per GPU. It takes three primitives, i.e., $R_0$, $R_1$, and $R_2$, to reduce the partial data with length $d$ on each GPU.

Single-step collective design may yield redundant data movement. For example, the All-reduce in Table~\ref{tab:composition} (Single) is not efficient because the total data moved is $dp^2$. This problem can be solved by formulating All-reduce as Reduce-scatter followed by an All-gather, which moves data of size $dp$ (see Figure~\ref{fig:allreduce}). The problem is that the All-gather depends on the result of the Reduce-scatter, violating the single-step design principle. We next describe the design of such multi-step collectives.

% \usepackage{multirow}
% \usepackage{graphicx}
\begin{table}[t]
\centering

\caption{Composition of Collective Functions on $p$ Processes}
\label{tab:composition}
\resizebox{0.9\columnwidth}{!}{
  \begin{threeparttable}
\begin{tabular}{lll}
\textbf{\# Steps}              & \textbf{Collective}        & \textbf{Composition}              \\ \hline
\multirow{8}{*}{Single}
                           & Broadcast                  & $M(i,\boldsymbol{U},dp) $           \\
                           & Reduce                     & $R(\boldsymbol{U},j,dp,\textit{op})$            \\
                           & All-gather                 & $\sum_iM(i,\boldsymbol{U},d)$       \\
                           & Reduce-scatter             & $\sum_jR(\boldsymbol{U},j,d,\textit{op})$       \\
                           & All-reduce                 & $\sum_jR(\boldsymbol{U},j,dp,\textit{op})$      \\
                           & Scatter                   & $\sum_jR(i,j,d,\textit{op})$          \\
                           & Gather                    & $\sum_iM(i,j,d)$ \\
                           & All-to-all                & $\sum_i \sum_jM(i,j,d)$ \\ \hline
\multirow{6}{*}{Multiple} & Broadcast                  &  All-gather $\cdot$ Scatter \\
                           & Reduce                     & Gather $\cdot$ Reduce-scatter \\
                           & All-gather                 & Broadcast $\cdot$ Gather  \\
                           & Reduce-scatter             & Scatter $\cdot$ Reduce \\ \cline{2-3}
                           & \multirow{2}{*}{All-reduce} & Broadcast $\cdot$ Reduce  \\
                           &                            & All-gather $\cdot$ Reduce-scatter                
\end{tabular}%
    %\begin{tablenotes}
    %\item Multi-step composition should be read from right-to-left, where $\cdot$ represents the fence primitive.
    %  \footnotesize
    %  \item *Summations ($\sum$) are in range $(0,p-1)$.
    %  \item ${}^\dagger$Composed of point-to-point ($i$-to-$j$) communications.
    %\end{tablenotes}
  \end{threeparttable}
}
\end{table}

\subsection{Multi-Step Collectives}\label{sec:composite}

A multi-step collective is a sequence of single-step collectives, where each step depends on the previous step.  For composing multi-step collectives, \codename{} exposes a \emph{fence} to express data dependencies. Table~\ref{tab:composition} (Multiple) shows example formulations of composite collectives as a sequence of two single-step collectives. In this algebraic formulation, the order of operations is from right to left, where $\cdot$ represents the fence between the operations. %Sometimes multi-step collectives yield a higher throughput than single-step collectives.

\begin{figure}[t]
\includegraphics[width=\columnwidth]{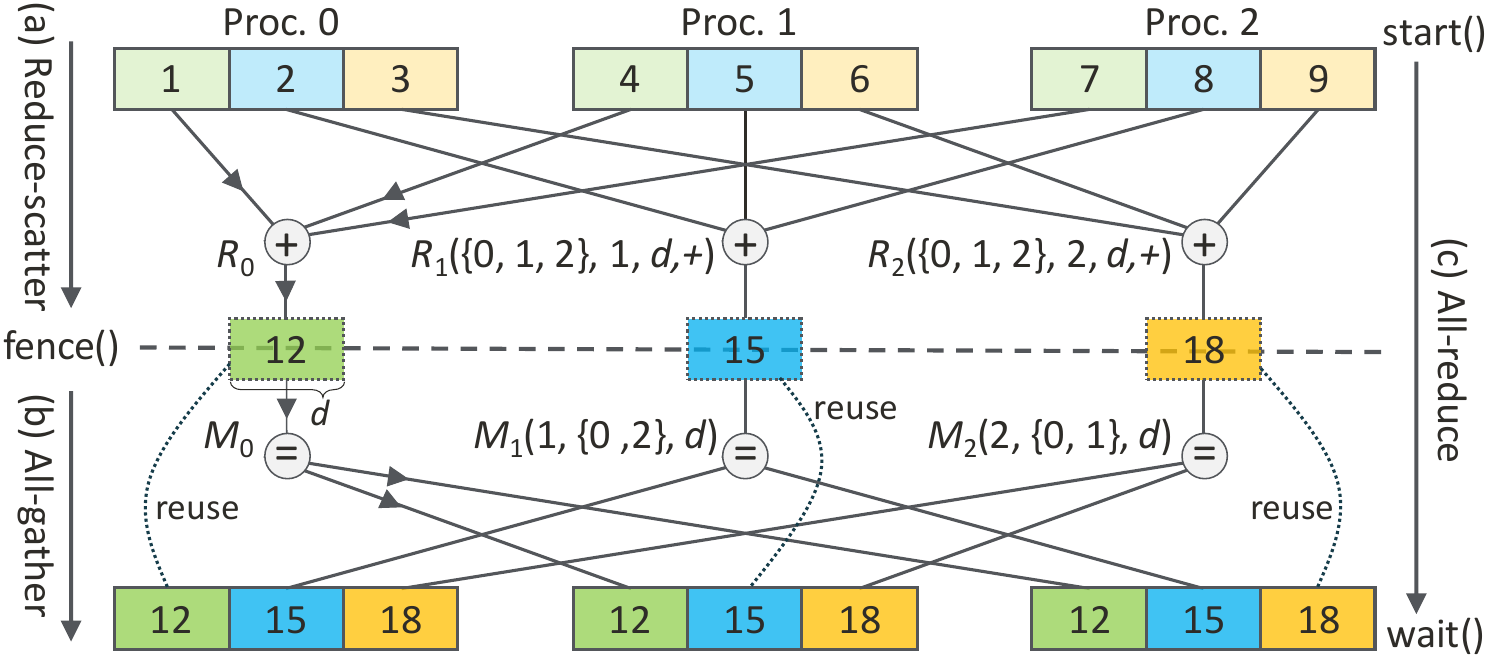}
\centering 
\caption{Composition of (c) All-Reduce function as (a) Reduce-Scatter followed by an (b) All-Gather on three processes. The registration takes three reduction primitives, followed by a fence, and then followed by three multicast primitives. The dashed edges on the broadcasts can be omitted for in-place implementation.}
\label{fig:allreduce}
\Description[<short description>]{<long description>}
\end{figure}

We use All-reduce\footnote{All-reduce performance is critical in scientific simulation and machine learning applications.} as a motivating example because its multi-step form is more efficient than the single-step form. Figure~\ref{fig:allreduce} shows the composition of a Reduce-scatter followed by an All-gather that is functionally equivalent to the single-step All-reduce, but has higher  throughput. In the multi-step algorithm, each GPU first reduces partial data from all GPUs and then broadcasts the result. To build this pattern, the user registers 1) $p$ reduction primitives, 2) a fence, and 3) $p$ multicast primitives in sequence, as shown in Listing~\ref{listing:allreduce} Lines, 4--11, where \texttt{all} is the vector $\{0, 1, 2, \cdots, p-1\}$.   Note that the receive buffer is reused for reducing partial data and therefore \texttt{others} represents the vector of all ranks but that of the root GPU.

%, xleftmargin=3.0ex]
\begin{lstlisting}[caption={Sample program for composing All-reduce as a Reduce-scatter followed by an All-gather. Lines 4--11 shows the composition step (Section~\ref{sec:design}) and Lines 13--17 shows the runtime optimization parameters (Section~\ref{sec:optimization}).}, label=listing:allreduce, float=t]
using namespace HiCCL
// persistent communicator
Comm<float> comm();
// step 1) register Reduce-scatter using primitives
for(int j = 0; j < numproc; j++)
  comm.add_reduction(sendbuf + j * count, recvbuf + j * count, count, all, j, op::sum);
// step 2) register fence to express data dependency
comm.add_fence();
// step 3) register All-gather using primitives
for(int i = 0; i < numproc; i++)
  comm.add_multicast(recvbuf + i * count, recvbuf + i * count, count, i, others);
// optimization parameters for Aurora
std::vector<int> hierarchy{numproc / 12, 6, 2};
std::vector<Library> library{MPI, IPC, IPC};
int stripe(1); // number of stripes, default: 1
int ring(1); // node count in the ring, default: 1
int pipeline(16); // pipeline depth, default: 1
// initialization
comm.init(hierarchy, library, ring, stripe, pipeline);

comm.start(); // nonblocking start
// do other things...
comm.wait(); // blocking wait

\end{lstlisting}

The fence is not a barrier, but a mechanism to express data dependencies between collections of primitives. \codename{} enforces the fine-grain dependencies between individual primitives in different steps without a barrier. For example, in Figure~\ref{fig:allreduce}, $M_0$ depends on $R_0$, $M_1$ depends on $R_1$, and $M_3$ depends on $R_3$. During execution, $M_0$ executes immediately after $R_0$'s completion, independent of $M_1$ and $R_2$. However, it is inefficient for $M_0$ to wait until the output of $R_0$ is complete. Pipelined execution (Section~\ref{sec:pipelining}) solves this problem by overlapping the execution of all primitives (on different data) without violating data dependencies that are expressed with the logical fences.

{
%\color{gray}
Listing~\ref{listing:allreduce} presents registration of primitives into the communicator created in Line 3. Once the composition of primitives is registered, the communicator is initialized in line 19 of Listing~\ref{listing:allreduce} with the optimization parameters (Section~\ref{sec:machine}). The \codename{} design offers \texttt{start()} and \texttt{wait()} functions for running the collective function. The former initiates the optimized communications from the CPU (although they take place on GPUs) and returns immediately. The latter blocks the CPU until the communication buffers on the corresponding GPU are safe to be reused. 
%In application code, user can overlap other operations between the \texttt{start()} and \texttt{wait()} functions.
}

\section{Optimizations}\label{sec:optimization}
\codename{} applies a set of hierarchical optimizations to any collective pattern built with the  primitives in Section~\ref{sec:primitives}. These optimizations depend on a set of machine-specific parameters that are explained in this section. %The optimizations are applied to an abstract machine that is expressed by the user through the parameters we explain in this section. Thus synthesizing the optimal communication pattern. %We also need to describe a conceptual machine for \codename{} create and implements the optimal pattern efficiently on the physical machine.

\subsection{Optimization Space}\label{sec:machine}

%{\color{gray} \codename{} applies two main optimizations: 1) factorization of primitives, and 2) pipelining of the collective data movement.}
%The end-to-end execution of a collective function is optimized if the template machine matches the physical machine as much as possible. Therefore we designed an abstract machine considering a few types of contemporary multi-GPU, multi-NIC node architectures, including exascale machines from different vendors.
HiCCL's optimization space is described with five parameters:
\begin{enumerate}
    \item Integer factors of $p$ for describing the network hierarchy (a vector).
    \item The choice of implementation library for point-to-point communication for each level (a vector).
    \item The striping factor for NICs ($s$).
    \item The number of nodes for a ring ($n$).
    \item Pipelining depth ($m$).
\end{enumerate}
The parameters 1, 3, and 4 depend on the machine architecture, while 2 depends on the communication software stack and 5 depends on the message length.
Note that HiCCL does not provide its own point-to-point communication operations, rather HiCCL can leverage the best available operations, including using different libraries for different levels of the communication hierarchy.
%and are used for optimization and execution of the primitive pattern. The parameters are

The optimizations are applied at the initialization step, after the collective composition (Section~\ref{sec:design}) is defined.
%In the initialization step, all the parameters in the optimization space must be specified.
As an example, Listing~\ref{listing:allreduce} shows the parameters for a specific system in Lines 12--17. \codename{} does not automatically select these parameters, which are part of the input; however, in our experiments we found that we were able to reuse the same description of the network hierarchy for all collective communication operations on a particular machine.  The parameters represent a virtual communication hierarchy that need not match the physical communication hierarchy, but of course the best performance will be achieved when the specified hierarchy matches  the underlying machine. %The parameters can be found by a generic auto-tuner [cite ATF] or by the user in a few educated attempts. The rest of this section explains the optimization parameters with greater detail.
%\codename{} then breaks down each primitive recursively into a directed asyclic graph (DAG) while preserving the original data dependencies between the root and the leaves (recall Figure~\ref{fig:primitives}). For multi-step collectives, the dependency is kept across fences. The structure of the DAG depends on the original primitive pattern, and the optimization heuristics that we describe in the rest of this section.

%\subsection{Hierarchical Optimization}\label{sec:hierarchical_factorization}

\begin{figure}[t]
\includegraphics[width=\columnwidth]{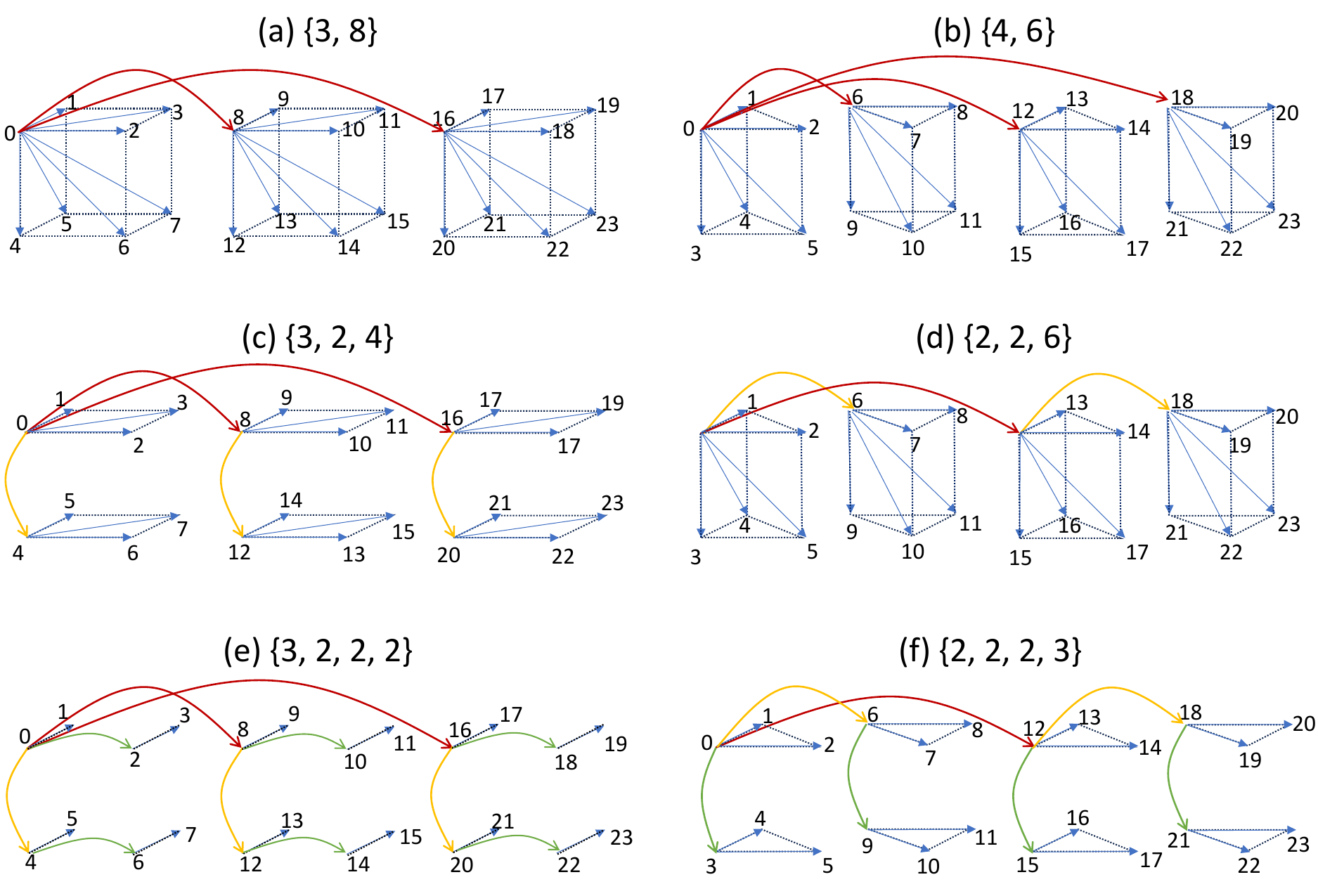}
\centering 
\caption{Various tree structures and their notations across 24 GPUs. The examples shows (a)--(b) two, (c)--(d) three, and (e)--(f) four levels of hierarchies.  The colors represents different communication links across: level 1 (red), level 2 (yellow), level 3 (green), and leaf (blue) levels. \codename{} implements each level with a the chosen communication library.}
\label{fig:machine_template}
\Description[<short description>]{<long description>}
\end{figure}
 
\subsection{Hierarchical Tree Structure}\label{sec:tree}

To exploit different network architectures, we parameterize the shape of the network hierarchy (see Figure~\ref{fig:machine_template}) using a vector of integer factors (Listing~\ref{listing:allreduce}, Line 13) that specifies a multilevel communication tree.
Figure~\ref{fig:machine_template}~(e) $\{2, 6, 2\}$ represents a three-level tree structure of 24 GPU endpoints. Assume there are two nodes and each node involves six devices, and each device involves two GPU dies. At Level 3, the first factor 2 partitions the endpoints into two groups of 12 At level 2, the two nodes are sub-divided into six groups of two dies each. The leaf level 1 specifies that the two dies within a device form a single group.  

%There are 20 possible factorizations (of 24) that can be used for forming balanced tree structures. See Figure~\ref{fig:machine_template} for other examples. The one-factor~[Bruce] (direct) case, denoted as $\{24\}$, is directly implemented across 24 GPUs. A prime factorization yields a four-level structure depending on the order of prime factors, as shown in Figure~\ref{fig:machine_template}~(e)--(f). For effective optimization, the tree must match to the physical communication hierarchy.

HiCCL assumes that the rank of each process/GPU is assigned in a way that reflects the network hierarchy; for example, Figure~\ref{fig:machine_template}(c) represents a machine with 4 GPUs per node. Every sequence of four rank ids, where the first is divisible by 4, represents the GPUs in a single node.  Thus, the desired grouping of GPUs at each level is fully determined by the machine description and the numbering of ranks.
The examples in Figure~\ref{fig:machine_template} are shown for a full set of leaves (GPUs). 
In case of custom collectives, the tree structure is pruned according to the sparsity of the leaf GPUs.

%The associated processes are assumed to be assigned in the physical machine with the pattern as sketched in the figure. In other words, consecutive ranks are assigned to each group of processors with the closest affinity at each scale of the physical network hierarchy. In this way, the hierarchical communication (Figure~\ref{fig:hierarchical_broadcast}) is applied automatically for the multicast and reduction primitives for an arbitrary hierarchy.

In practice, each level of the communication hierarchy has separate hardware and software, which means that different libraries may have very different performance at different levels of the hierarchy. As noted above, \codename{} allows specific libraries to be used at specific levels of the hierarchy to  optimize communication for a specific system. For the mixed-library implementation of \codename{}, the library used for communication at each level of the hierarchy is specified as in Listing~\ref{listing:allreduce}, Line 14. See Section~\ref{sec:portability} for options.

%{
%\color{gray}
%\textbf{Point-to-point:} The tree heuristic solves this problem automatically for Scatter and Gather, where there are $p$ point-to-point (See Table~\ref{tab:composition}) communications of $d$ bytes.The tree heuristic utilizes all NICs automatically except. The root node stages each point-to-point communication on a different GPU in the root node so that all NICs are utilized. This fact does not hold for broadcast ($B(i,\overline{j},d)$) and reduce ($R(\overline{i},j,d)$) because they are atomic primitives and their data ($d$) that cannot be divided automatically. For enforcing striping for the primitives, we implemented the striping heuristic.
%}

\subsection{Multi-NIC Striping}\label{sec:striping_optimization}
\begin{figure}[t]
\includegraphics[width=\columnwidth]{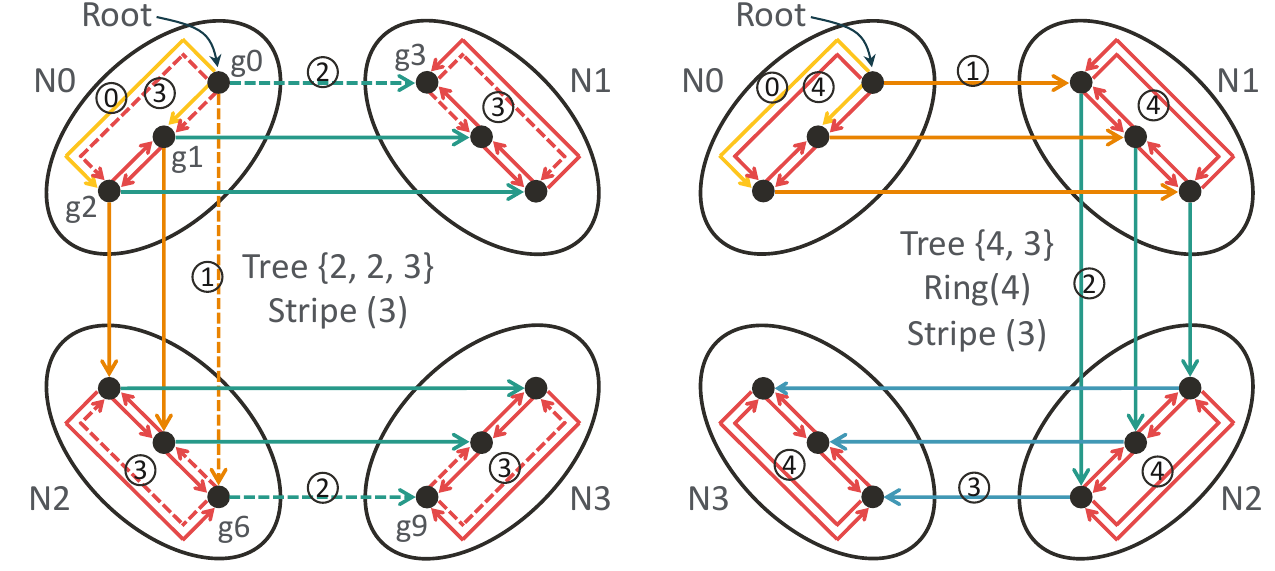}\\
\footnotesize \hspace{0.1cm} (a) Tree \hspace{3.6cm}(b) Ring
\centering 

\caption{(a) Tree and (b) ring factorizations on four nodes with three GPUs each. Striping forms multi-rail patterns that utilizes all GPUs (and hence all NICs). With striping, (a) tree forms four stages \textcircled{0}--\textcircled{3} and (b) ring forms five stages \textcircled{0}--\textcircled{4}, where each stage depends on the previous. The dashed route in (a) tree is shown for a single stripe originating from the root.}
\label{fig:tree_ring_template}
\Description[<short description>]{<long description>}
\end{figure}

Consider a broadcast factorized on 12 GPUs as $\{2,2,3\}$. The last level represents a node with three GPUs each. The resulting tree structure has three levels, which are shown with dashed lines in Figure~\ref{fig:tree_ring_template}~(a). The message hops across nodes as \textcircled{1} orange dashed hop (g0-to-g6) and \textcircled{2} green dashed hops (g0-to-g3 and g6-to-g9). Then each staging GPU (g0, g3, g6, g9) multicasts the data within nodes with \textcircled{3} red dashed hops.
The problem is that each staging GPU is potentially bound to a single NIC (Section~\ref{sec:multi_NIC}), and therefore the hops across nodes underutilize the multi-NIC node architecture.

The aim of striping is forming multi-rail pattern by employing all GPUs (and hence all NICs) in the inter-node communication (Section~\ref{sec:multi_NIC}). This heuristic ``stripes'' each primitive into $s=3$ parallel branches in the corresponding root node, since there are $g=3$ GPUs per node. This is achieved by an additional set of intra-node communication, as shown with \textcircled{0} solid golden hops. Striping is internally composed as $\sum_j^{\textrm{\{g1, g2\}}}R(0,j,d/s)$, followed by a fence (Section~\ref{sec:composite}). For preserving the original dependencies, each GPU in the root node (g0, g1, g2) must multicast the corresponding partial data to all GPUs, i.e., $\sum_i^{\{\textrm{g0, g1, g2}\}}M(i,\texttt{others},d/s)$, where $\texttt{others}$ represent all GPUs but $i$. The three parallel branches are further factorized recursively down to the point-to-point dependencies. As a result, striped factorization produces multi-rail communication patterns, automatically utilizing all NICs in the participating nodes. Figure~\ref{fig:tree_ring_template} shows striping of (a) tree (making all striped lines solid) and (b) hybrid ring+tree (explained next).

\subsection{Hybrid Ring+Tree Topology}\label{sec:ring}

A ring forms a communication chain across $n$ conceptual nodes. Figure~\ref{fig:tree_ring_template}~(b) gives an example for \texttt{ring(4)}, where \texttt{4} represents the number of nodes. To fully utilize the NICs, we employ striping by setting \texttt{stripe(3)} because there are three GPUs per node, resulting in the extra \textcircled{0} golden hops. The striped ring starts from the root node and each stripe unfolds across \textcircled{1} (N0 to N1), \textcircled{2} (N1 to N2), and \textcircled{3} (N2 to N3), until terminating in the last leaf node (N3). Finally, the partial data is assembled on all leaf GPUs with \textcircled{4} red hops, that are factorized using a tree within each node, resulting a hybrid ring+tree pattern. A tree-only communication pattern is achieved using \texttt{ring(1)}, where there is only one conceptual node with all ($p$) GPUs and effectively no ring.
%, i.e., $\sum_{n=0}^4\sum_{i=0}^3B(3n+i,\{3n, 3n+1, 3n+2\}, d/3)$.

Overall, \codename{} factorizes each primitive with 1) striping, 2) ring, and 3) tree (in this order)---down to a dependency graph composed of multiple point-to-point communication stages. In the execution of this graph, each stage depends on the previous one, causing idle GPU time. To minimize idle time, \codename{}  overlaps the communication stages with a generalized pipeline as we explain next.

%For example, Figure~\ref{fig:tree_ring_template} shows the factorization of $B(0,\texttt{all}, d)$ with (a) striped tree and (b) striped ring across 12 GPUs with four nodes. The (a) tree is not balanced and involves a bottleneck at the root node. The root node sends out six stripes of $d/3$, a total of $2d$ bytes. On the other hand, each node in the (b) ring is a jumping point of $d$ data across nodes and therefore balanced. Nevertheless, when the data is completely staged, the cost of ring is high---$\sim O(n)$. On the other hand the complexity of tree across nodes is $O(\log n)$. The ring is 

%and theoretically is two times faster than the (a) tree.

\subsection{Pipelining}\label{sec:pipelining}

\begin{figure}[t]
\includegraphics[width=\columnwidth]{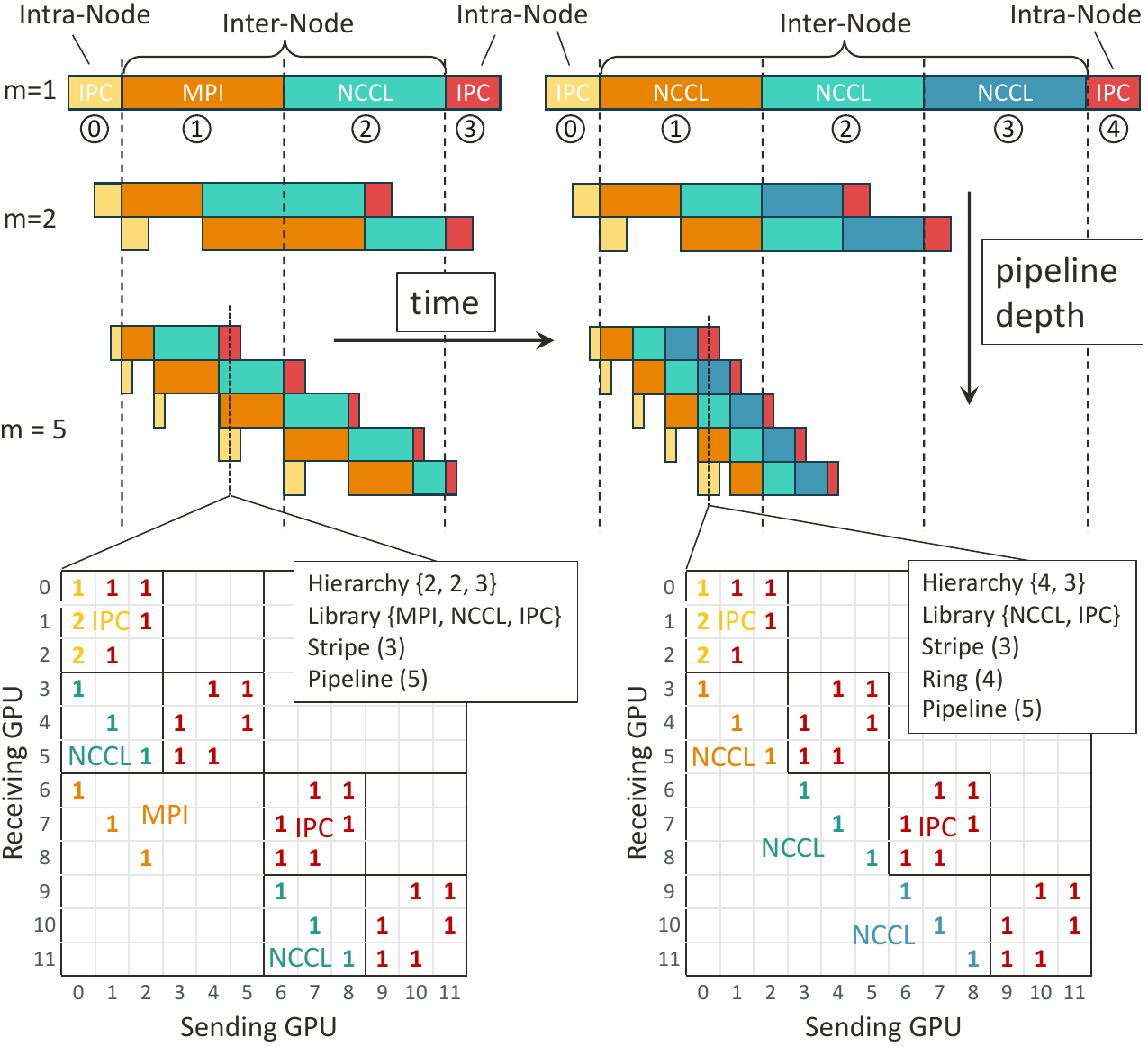}
\footnotesize \hspace{0.1cm} (a) Tree \hspace{3.6cm}(b) Ring
\centering
\caption{Pipelining and the overlapped  pattern of hierarchical communications for broadcast with (a) tree and (b) ring+tree virtual topologies. The machine model that produces these patterns are noted in the figure. In this case, \codename{} employs different implementation libraries (specified---MPI, NCCL, or IPC) for mixed communications across levels of the hierarchy.}
\label{fig:pipelining}
\Description[<short description>]{<long description>}
\end{figure}

HiCCL employs a pipelining optimization that overlaps communication across all steps of a multi-step collective operation. A sequential execution pipeline is made up of a series of stages that are marked with numbers as shown in Figure~\ref{fig:pipelining} ($m=1$), where $m$ is the pipeline depth. Each stage involves 1) a collection of point-to-point communications, or 2) the former followed by a set of computations (if a reduction is involved) to be executed on GPUs.%with different colors and marked with numbers, e.g., \textcircled{0} (gold), \textcircled{1} (orange), etc. A sequential pipeline has a single channel  and executes a single stage at a time.  %This pipeline design is not specific to a particular network or collective, but for the correct result, the units must be subsequently executed in their original (sequential) order.

Figure~\ref{fig:pipelining}(a) shows the pipeline for the tree example in Figure~\ref{fig:tree_ring_template}(a). When $m=1$, stage \textcircled{0} corresponds to the intra-node  striping, \textcircled{1}--\textcircled{2} corresponds to a two-level binary tree across nodes, and \textcircled{3} corresponds to the intra-node assembly. Similarly, Figure~\ref{fig:pipelining}(b) shows the pipeline for the ring example in Figure~\ref{fig:tree_ring_template}(b). The original ring pipeline ($m=1$) requires three stages across nodes that takes approximately 1.5 times longer to complete. By overlapping communications across nodes the ring  is two times faster than a tree for this example.

To overlap the execution of multiple stages, without violating data dependencies, we partition the original payload ($d/s$) bytes across $m$ {\em channels} in the pipeline, where each channel executes on $d/s/m$ bytes at a time. We overlap all channels in the final form of the execution pipeline as seen in Figure~\ref{fig:pipelining} ($m=5)$. In the final form, there are a few partially overlapped stages in the ``warm-up'' and ``wind-down'' of the pipeline, whereas the middle stages are fully overlapped. To overlap all stages, the pipeline must be deeper than the number of stages, which requires at least four channels for (a) tree and five channels for (b) ring in the examples in Figures~\ref{fig:tree_ring_template}--\ref{fig:pipelining}.

%All GPUs execute the same pipeline structure in parallel. The dependencies across GPUs are sparse and complex. A pipeline stalls when the associated GPU waits on a completion of a step, where another GPU with no dependency moves on executing subsequent steps. 

%Overlapping provides speedup only if data movement utilize different resources. For example, communications across nodes and within nodes can be completely overlapped.

The overlapped communication pattern changes across the stages of the pipeline. To illustrate, we represent the fully overlapped pattern with a hierarchical communication matrix in Figure~\ref{fig:pipelining} (bottom) for the (a) tree and (b) ring examples. Each non-zero entry in the matrices corresponds to a point-to-point communication. All entries are executed with the specified library for that level of the hierarchy. For demonstrating the mixed use of libraries, (a) tree uses \textcircled{1} MPI across the two groups of 6 GPUs, \textcircled{2} NCCL across nodes in the same group, and \textcircled{3} IPC within nodes, i.e., $3\times 3$ diagonal blocks in the matrices in Figure~\ref{fig:pipelining} (bottom). In (b) ring, we assume no hierarchy across nodes, and we use IPC in the diagonal blocks and NCCL elsewhere.

%{
%\color{gray}
%\textbf{Fusing Computations:} The reduction heuristics produces additional communications followed by computations. We ``fuse'' them into a single command by scheduling GPU asynchronously (works for NCCL and IPC). In this way, we hide the GPU kernel call latency and minimize the wait time for command completion, and hence the stalls on the distributed asynchronous pipeline.
%}

\subsection{Performance Model}\label{sec:model}

The communication cost for the Broadcast examples with pipelining are shown for \eqref{eq:ring} ring and \eqref{eq:tree} tree on $n$ conceptual nodes with no physical hierarchy. The first and second terms correspond to inter-node and intra-node communication cost, respectively. Here, the variables represent $\alpha$: communication latency, $m$: pipeline depth, $k$: number of NICs (per node), $f$: NIC bandwidth, and $d$: message length.

\begin{equation}\label{eq:ring}
    t_\textrm{ring} = \left(\alpha+\frac{d}{kfm}\right)(n+m-2) + \mathcal{O}(d/m)
\end{equation}
\begin{equation}\label{eq:tree}
    t_\textrm{tree} = \left(\alpha m+\frac{d}{kf}\right)\log n + \mathcal{O}(d/m)
\end{equation}

We model the intra-node cost as roughly $\mathcal{O}(d/m)$ because the details depend on the specific node architecture. Nevertheless, pipelining hides that cost and the residual intra-node overhead shrinks with $1/m$, as clearly seen in Figure~\ref{fig:pipelining} (red stages). Asymptotically, an infinitely deep pipeline ($m\rightarrow \infty)$ with zero latency ($\alpha=0$) hides the intra-node communication overhead completely. In practice,  pipelining is most useful for large message sizes, where the latency term is negligible. Asymptotic pipelining yields $t_\textrm{ring}\sim d/kf$ that does not depend on the number of nodes, i.e., $\mathcal{O}(1)$. On the other hand, $t_\textrm{tree}\sim d\log n /kf= \mathcal{O}(\log n)$. On four nodes ($n=4$) ring is theoretically two times faster than tree as we will show experimentally in Section~\ref{sec:primitive_performance}.

% \usepackage{multirow}
\begin{table}[h]
\centering
\caption{Asymptotic Collective Throughput in GB/s}
\label{tab:upper_bounds}
\footnotesize

  \begin{threeparttable}
\begin{tabular}{cccc}
\textit{Broadcast} & \textit{Gather / All-Gather}      & \multirow{2}{*}{\textit{All-Reduce}} & \multirow{2}{*}{\textit{All-to-All}} \\
\textit{Reduce}    & \textit{Scatter / Reduce-Scatter} &                                      &                                      \\
$kf$               & $kf\dfrac{p}{p-g}$                & $kf\dfrac{p}{2(p-g)}$                & $kf\dfrac{p}{g(p-g)}$ 

\end{tabular}

\begin{tablenotes}
\item $p$: total \# participated GPUs, $g$: \# GPUs per node, $k$: \# NICs per node, and $f$: rated NIC bandwidth in GB/s.
\end{tablenotes}

\end{threeparttable}
\end{table}

As our theoretical baseline of all collectives in Table~\ref{tab:collectives}, we consider the upper bounds for throughput (GB/s) as shown in Table~\ref{tab:upper_bounds}. This upper bound solely depends on the total number of participating GPUs, collective pattern, the number of NICs and GPUs per node, and the rated bandwidth per NIC. %The idea behind \codename{}'s pipelining optimization is hiding the intra-node cost (which can be complicated to model), and hence converging to the hierarchical network limits with the theoretical NIC bandwidth.

\section{Implementation}
\label{sec:impl}

HiCCL is implemented in C++ for distributed applications, libraries, and frameworks running on GPUs (and CPUs). HiCCL requires MPI for setting up.

% \usepackage{graphicx}
%\begin{table}[t]
%\centering
%\caption{Compiler flags (columns) and integrated libraries (rows) for porting across systems of different hardware.}
%\label{tab:port_libraries}
%\small
%\resizebox{\columnwidth}{!}{%
%\begin{tabular}{l|cccc}
% &  CPU (D)     & \texttt{PORT\_CUDA}        & \texttt{PORT\_HIP}        & \texttt{PORT\_SYCL}      \\\hline
%\texttt{MPI}            &  MPI     & GPU-MPI    & GPU-MPI    & GPU-MPI   \\
%\texttt{XCCL}           &              & NCCL              &  RCCL          &                 \\
%\texttt{IPC}            &              & CUDA IPC & HIP IPC &  L0 IPC \\       
%\end{tabular}%
%}
%\end{table}

\subsection{Implementation Options} \label{sec:portability}
We implemented \codename{} by leveraging the point-to-point communication API of the chosen library for each level of the hierarchy. We integrated non-blocking point-to-point functions of MPI, NCCL, RCCL, and OneCCL to be used within and across nodes, and vendor-provided (CUDA, HIP, or Level Zero) IPC put \& get to be used within nodes.  We also integrated CUDA, HIP, and SYCL programming models for targeting Nvidia, AMD, and Intel GPUs, respectively.% The combination of choices are shown in Table~\ref{tab:port_libraries}.

\subsection{Persistent Design}

\codename{} takes advantage of repetitive collective function calls by memoizing the optimized data movement and scheduling in internal data structures. On the second and subsequent uses of a communication operation, these structures are reused to avoid the cost of making on-the-fly decisions. Furthermore, \codename{} involve no global synchronization either in the composition and synthesis nor in the execution. %The remaining inherent overhead is a few function calls on the order of 35--130 nanoseconds which is insignificant compared to network latency (2--200 microseconds). 

%\subsection{Memory Management}

%The user is responsible to allocate the original communication buffers and register them through the primitives in the composition step. \codename{} internally allocates memory on CPU and GPU. The CPU memory is used for metadata (KB-scale), and therefore is not significant. The GPU memory is used for staging communication data, which can be significant. To minimize the memory footprint, \codename{} reuses the original buffer space that is provided by the user and recycles its own memory with respect to the dependencies in the execution.

%\subsection{Interoperability}\label{sec:interoperability}

%\codename{} is implemented for inter-operating with other libraries, applications, and runtimes. We already made \codename{} work with a non-MPI runtime tool [FlexFlow] for accelerating the All-reduce in a machine learning application on Frontier system, where RCCL and MPI performance (evidently in Figure~\ref{fig:overview_bandwidth}~(c)) is not satisfactory.

\begin{figure*}[t]
\includegraphics[width=1.015\textwidth]{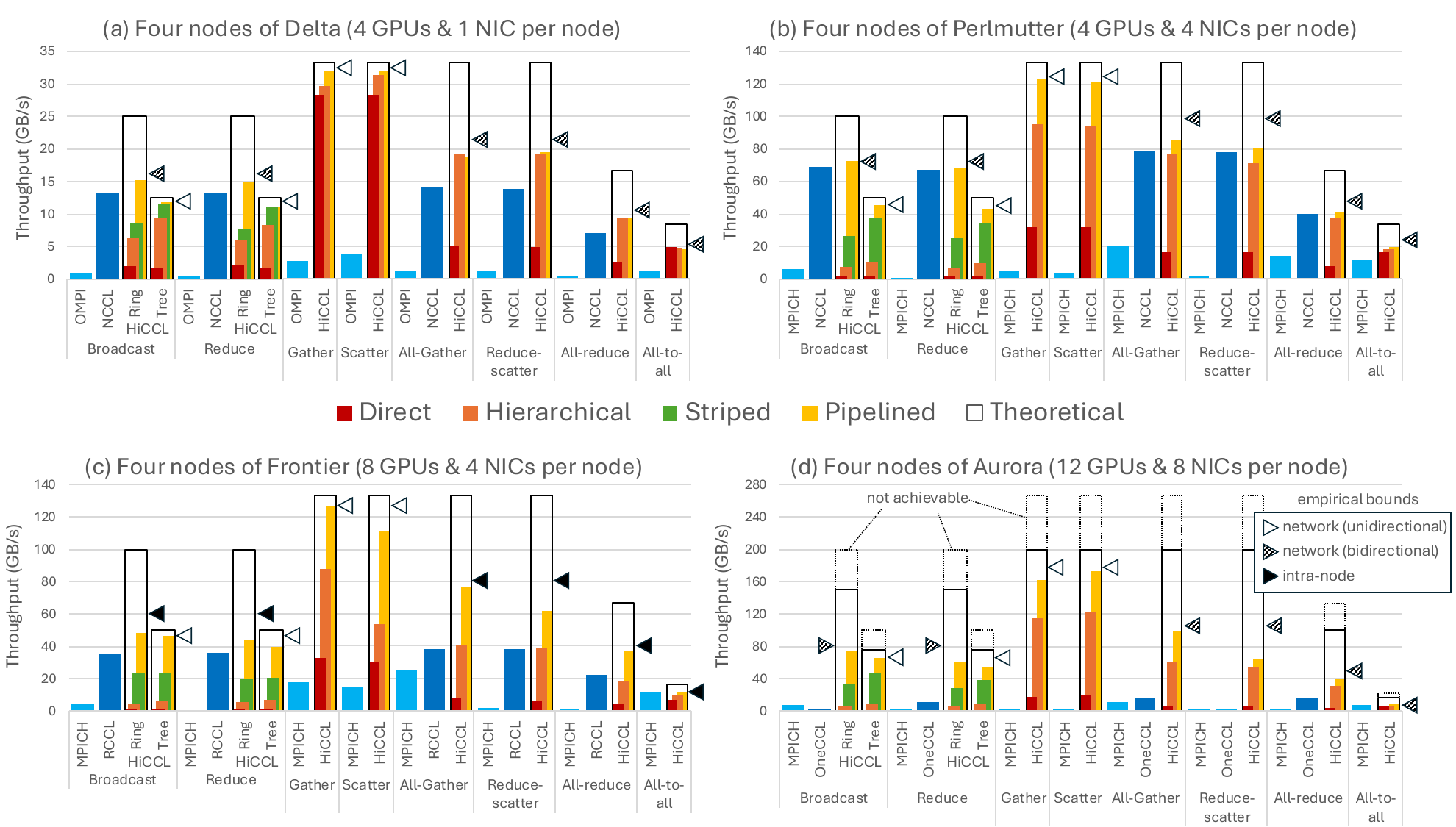}
\centering 
\caption{Peak collective throughput (GB/s) on (a) Delta, (b) Perlmutter, (c) Frontier, and (d) Aurora. HiCCL optimizations are applied incrementally, where the frames around bars represents the theoretical limit based on the rated NIC performance. Triangle marks represents empirical bounds based on isolated measurements across two nodes. HiCCL throughput reaches to the empirical bounds on all systems, demonstrating performance portability across architectures.}
\label{fig:hiccl_collectives}
\Description[<short description>]{<long description>}
\end{figure*}

\section{Evaluation}\label{sec:evaluation}

We evaluate the performance portability of \codename{} across eight commonly used collective functions  in Table~\ref{tab:collectives} and four HPC systems with different hardware and software.

\subsection{Experimental Setup}

The node architectures of the test systems are summarized in Table~\ref{tab:system_summary}, showing various numbers of CPUs, GPUs and NICs per node. We compare \codename{} collective throughput with that of a) corresponding native MPI implementations and b) vendor-provided collective communication libraries (NCCL, RCCL, or OneCCL) on each system. The MPI implementations are based on OpenMPI (OMPI) for Delta and modified versions of MPICH for Perlmutter, Frontier, and Sunspot. NCCL and RCCL uses AWS-ofi extension for portability to Slingshot (SS-11) networks that the systems incorporate~\cite{aws_ofi}. %As our baseline, we use the MPI and NCCL collective functions in Table~\ref{tab:collectives}.

% \usepackage{graphicx}
\begin{table}[t]
\caption{Summary of Node Architecture of Test Systems}
\label{tab:system_summary}
\centering
\resizebox{\columnwidth}{!}{%
\begin{threeparttable}
\begin{tabular}{lllll}
\textbf{System} & \textbf{CPUs} & \textbf{GPUs} & \textbf{NICs} & \textbf{B/W}${}^\dagger$\\ \hline
Delta           & 1 AMD EPYC          & 4 Nvidia A100       & 1 SS-11 & 25 GB/s\\
Perlmutter      & 1 AMD EPYC          & 4 Nvidia A100       & 4 SS-11 & 100 GB/s\\
Frontier        & 1 AMD EPYC          & 8 AMD MI250x*       & 4 SS-11 & 100 GB/s\\
%DGX-A100        & 2 AMD EPYC          & 8 Nvidia A100       & 8 IB     & 200 GB/s\\
Aurora         & 2 Intel Xeon        & 12 Intel PVC*         & 8 SS-11 & 200 GB/s\\
\end{tabular}
    \begin{tablenotes}
      \item *Each AMD MI250x and Intel PVC device involves two processor ``dies'' or ``tiles'', which we count as separate GPUs. 
      %\item *Each AMD MI250x~\cite{pearson2023interconnect} and Intel PVC~\cite{moore_2022} involves two processor dies, which we count as separate GPUs. 
      \item ${}^\dagger$ Rated unidirectional node bandwidth based on the number of NICs.
    \end{tablenotes}
\end{threeparttable}
}
\end{table}

\subsection{Measurements}

We measure the peak throughput of each collective function on each system. We run the end-to-end collective function in multiple rounds: 5 warmup rounds and 10 measurement rounds. In each measurement round, we measure the elapsed time from a global synchronization to the moment that the communication buffers on all GPUs are safe to be reused. We run collectives with buffer sizes of $pd$ bytes. For example, Scatter sends $d$ bytes to each of the $p$ processors. If a collective requires $t$ second to execute, the  throughput is $dp/t$ (GB/s). We vary $d$ across large message sizes (larger than a MB) until the throughput saturates the achievable bandwidth.

\subsection{Collective Throughput}\label{sec:primitive_performance}

 Figure~\ref{fig:hiccl_collectives} (a)--(d) shows the peak collective throughput on four nodes of each test system. As an existing baseline, we use available library implementations as presented with light blue (MPI) and dark blue (NCCL/RCCL/OneCCL) colors. We confirmed these baseline results with administrators of each system, in some cases setting tuning flags when advised to do so. 

\subsubsection{Overall Speedup} In Figure~\ref{fig:hiccl_collectives}, HiCCL results are shown with four bars, representing the incremental effect of optimizations. When all optimizations are applied, HiCCL's geomean speedup over the MPI implementations is 12.52$\times$, 14.22$\times$, 9.76$\times$, and 48.02$\times$ on Delta, Perlmutter, Frontier, and Aurora, respectively. On the other hand, the speedup over vendor-provided libraries is 1.26$\times$, 1.05$\times$, 1.55$\times$, and 12.01$\times$ on the respective systems. The comparison suggests that while MPI implementations are not optimized for throughput, vendor-provided libraries generally are, with the exception of OneCCL on Aurora.

The rest of this subsection discusses the effect of individual optimizations of HiCCL on tested collective functions.

\subsubsection{Hierarchical Optimizations}
Red bars in Figure~\ref{fig:hiccl_collectives} represent direct implementations of collectives with non-blocking point-to-point functions, assuming there is no hierarchy across GPUs---i.e., the description of the network hierarchy for these experiments is just \{$p$\}, where $p$ is the number of participating GPUs. Direct implementations use NCCL on Delta and Perlmutter, and MPI on Frontier and Aurora as they are the most performant options. 

Orange bars in Figure~\ref{fig:hiccl_collectives} represent hierarchical optimizations with various factorizations (Section~\ref{sec:tree}) that are specific to each system. These factorizations are shown in the third column of Table~\ref{tab:experiment_parameters}, where the bold entries represent the hierarchies within nodes. Within nodes, Delta and Perlmutter are represented with a single level, e.g., \textbf{4}, due to the directly connected GPUs. On the other hand, Frontier and Aurora nodes consist two-level hierarchies, e.g., \{\textbf{4}, \textbf{2}\} and \{\textbf{6}, \textbf{2}\}, where the lower level represents the dual-die devices and the upper level represents high-bandwidth links betweeen these devices. Frontier has four devices and Aurora has six devices per node. On the other hand, the factorizations across nodes are to form virtual topologies with multi-rail tree (two levels) or ring (single-level) structures across nodes. Overall, hierarchical optimizations obtain a geometric average of 4.08$\times$ speedup over the direct baseline on all systems and collectives.

% \usepackage{multirow}
% \usepackage{graphicx}
\begin{table}[t]
\centering
\caption{Hierarchical Factorizations and Libraries Used in Figures~\ref{fig:hiccl_collectives}--\ref{fig:pipelining_result}}
\label{tab:experiment_parameters}
\footnotesize
%\resizebox{\columnwidth}{!}{%
\begin{tabular}{llll}
            \textbf{System} &  \textbf{Topology}  & \textbf{Hierarchy}      & \textbf{Implement. Library}                \\ \hline
Delta /  & Tree & \{2, 2, \textbf{4}\}    & \{NCCL, NCCL, \textbf{IPC}\}    \\
Perlmutter   & Ring+Tree & \{4, \textbf{4}\}       & \{NCCL, \textbf{IPC}\}          \\ \hline
\multirow{2}{*}{Frontier} & Tree & \{2, 2, \textbf{4}, \textbf{2}\} & \{MPI, MPI, \textbf{IPC}, \textbf{IPC}\} \\
                        & Ring+Tree & \{4, \textbf{4}, \textbf{2}\}    & \{MPI, \textbf{IPC}, \textbf{IPC}\} \\ \hline
%\multirow{2}{*}{DGX-A100} & Tree & \{2, 2, 8\} & \{NCCL, NCCL, IPC\} \\
%                        & Ring & \{4, 8\}    & \{NCCL, IPC\} \\ \hline    
\multirow{2}{*}{Aurora} & Tree & \{2, 2, \textbf{6}, \textbf{2}\} & \{MPI, MPI, \textbf{IPC}, \textbf{IPC}\} \\
                        & Ring+Tree & \{4, \textbf{6}, \textbf{2}\}    & \{MPI, \textbf{IPC}, \textbf{IPC}\} \\ \hline                        
\end{tabular}%
%}
%\\
%\footnotesize
%\raggedright
%*represents implementation with a get protocol and  protocol otherwise.
\end{table}

\subsubsection{Multi-NIC Striping} Green bars in Figure~\ref{fig:hiccl_collectives} represent the throughput with HiCCL's multi-NIC striping (Section~\ref{sec:striping_optimization}). This optimization is beneficial primarily for Broadcast and Reduce collectives, as these do not inherently utilize all NICs. On Delta, where each node has just one NIC, striping offers limited advantages, evidenced by a 1.29$\times$ speedup. This improvement is attributed to four GPUs utilizing the single NIC more effectively than would a solitary GPU. In contrast, on multi-NIC nodes, such as Perlmutter, Frontier, and Aurora, striping yields speedups of 3.62$\times$, 3.94$\times$, and 4.76$\times$, respectively, demonstrating significant throughput improvements.

\subsubsection{Pipelining} Yellow bars in Figure~\ref{fig:hiccl_collectives} represent the pipelining optimization (Section~\ref{sec:pipelining}).
% and ii) the theoretical limit (Section~\ref{sec:model}) of the each machine shown with hollow bars.
Pipelining hides intra-node communications with a  tree topology and also provides algorithmic speedup with ring+tree topology. To demonstrate, we test HiCCL's Broadcast and Reduce collectives with both virtual topologies, denoted as ``Tree'' and ``Ring'' in Figure~\ref{fig:hiccl_collectives}. We observe that the ring obtains up to 2.72$\times$ speedup (on Perlmutter) yet does not saturate the throughput up to its theoretical limit. All other collectives use the tree topology, and pipelining is effective on systems with significant intra-node communication yet obtains no more than two times speedup. To explain these limitations, we use empirical bounds that are explained next. 

%Overall, pipelining obtains an additional geomean speedup of 1.14$\times$ compared to no pipelining. Nevertheless, the collective throughput generally does not reach near to the theoretical limit as we will discuss next.

\subsubsection{Upper Bounds}\label{sec:evaluation_bounds}
The frames around our HiCCL results in Figure~\ref{fig:hiccl_collectives} represent the theoretical upper bounds that are given in Table~\ref{tab:upper_bounds}. 
Aurora is a special case because round-robin assignment (Section~\ref{sec:multi_NIC}) of 12 GPUs and  8 NICs. In this case, GPU $i$ is assigned to NIC $i\mod 8$, yielding GPUs 0--7 assigned to NICs 0--7 whereas GPUs 8--11 oversubscribe NICs 0--3.
%As a result, GPU 0 is assigned to NIC 0, GPU 1 is assigned to NIC 1, and so on until GPU 8 which is assigned to NIC 0 (oversubscribing it). The overall assignments of GPUs to NICs as ${0, 1, 2, 3,..., 7, 0, 1, 2, 3, 4}$
As a result, the first four NICs handle two GPUs each, whereas the remaining NICs handle a single GPU each, leading to load imbalance. Thus, the achievable bandwidth on Aurora is limited to 75\% of the theoretical bandwidth with this strategy.

Even though we apply all optimizations aggressively and use large buffer sizes to saturate throughput, we converged to only a fraction of the theoretical limits. %Theoretically (with an infinitely deep pipeline with zero latency), Equations (\ref{eq:ring})--(\ref{eq:tree}) suggest that ring will be two times faster than tree on four nodes. With \codename{} in Figure~\ref{fig:hiccl_collectives}~(a)--(b), ring is 1.94$\times$ and 1.55$\times$ faster than tree, respectively. Ring on (b) Perlmutter utilizes only 72\% of its theoretical limit.
To understand why, we measured the unidirectional and bidirectional bandwidth across two nodes in isolation rather than using the numbers in the spec sheet. These empirical upper bounds are indicated by hollow (unidirectional) and striped (bidirectional) triangles marks in Figure~\ref{fig:hiccl_collectives}.  For example, the empirical bounds for Gather and Scatter are considered unidirectional, as the bottleneck in these operations is the root node, which either receives or sends messages in only one direction. On the other hand, other collectives send and receive messages at the same time and therefore their empirical bounds are the bidirectional utilization.

A surprising result is that the intra-node communication cost on Frontier is higher than that of inter-node communication, which prevents us from hiding the former with the latter, even when we align the virtual hierarchy with the node architecture. Therefore we also measured the intra-node cost in isolation and marked the corresponding empirical bound with dark triangles in Figure~\ref{fig:hiccl_collectives}(c). Notwithstanding, our results suggest that \codename{} almost always comes close to the maximum potential of the machine in practice.

\begin{figure}[t]
\includegraphics[width=\columnwidth]{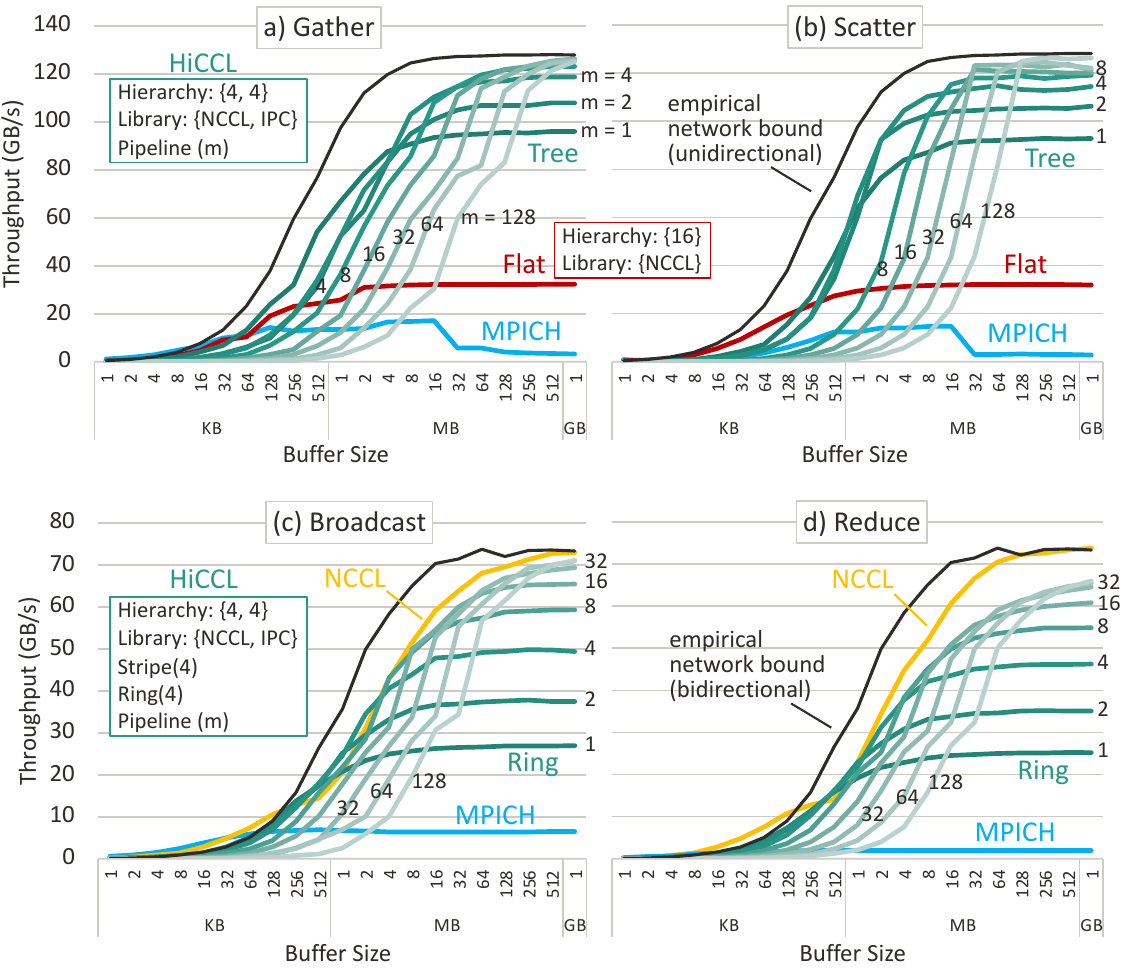}
\centering 
\caption{Throughput with various pipeline depths and buffer sizes for tree implementation of (a) Gather and (b) Scatter, and ring+tree implementation of (c) Broadcast and (d) Reduce on four nodes of Perlmutter.}
\label{fig:pipelining_result}
\Description[<short description>]{<long description>}
\end{figure}

%\textbf{Comparison with MPI and NCCL:} MPICH serves as Perlmutter's native MPI implementation. The blue curves in Figure~\ref{fig:pipelining_result} represent the MPI collective function performance (higher is better)\footnote{We confirmed the results in Figure~\ref{fig:pipelining_result} with MPICH developers.}. For (a) Gather and (b) Scatter, NCCL lacks the equivalents of \texttt{MPI\_Gather} and \texttt{MPI\_Scatter} functions as of version 2.1, and therefore we implemented them with point-to-point functions (\texttt{ncclSend} / \texttt{ncclRecv}) and labeled as Direct (maroon curves) in Figure~\ref{fig:pipelining_result}~(a)--(b).

\subsection{Pipeline Depth}

%Although pipelining improves the communication throughput, it must be used with taste. When used aggressively, message sizes becomes so small that the latency overhead between stages becomes dominant and hurts the overall throughput, as suggested in Equations \eqref{eq:tree}--\eqref{eq:ring}. For demonstrating this phenomenon, we vary the pipeline depth on four nodes of Perlmutter.

Although pipelining improves communication throughput, it must be used with caution. When applied too aggressively, message sizes become so small that the latency overhead between stages dominates, negatively impacting the overall throughput, as suggested in Equations \eqref{eq:tree}--\eqref{eq:ring}. To illustrate this phenomenon, we vary the pipeline depth across four nodes of Perlmutter

Figure~\ref{fig:pipelining_result} shows the performance curves of (a) Gather and (b) Scatter with tree, and (c) Broadcast and (d) Reduce with ring across various pipeline depths ($m$) 1 to 128, where $m=1$ means no pipelining and $m=128$ means pipelining with 128 channels. Pipelining clearly improves throughput for large message lengths. Nevertheless, excessive use for smaller messages reduces the throughput. 

%For (c) Broadcast and (d) Reduce, NCCL offers \texttt{ncclBroadcast} and \texttt{ncclReduce} functions whose performance is shown with the gold curves in Figure~\ref{fig:pipelining_result} (c)--(d), respectively. The results show that NCCL’s functions are highly optimized and have slightly better performance than \codename{} at some data points. NCCL throughput exceeds the empirical bound with small message sizes (e.g., 128 KB) because its virtual topology is adaptively changed to tree whereas the empirical bound is for ring that is efficient for large messages.

%See Figure~\ref{fig:pipelining_result}.

Further implications of pipelining depend on the optimizied communication pattern. In this case, we apply the tree algorithm for Gather / Scatter, and ring+tree algorithm for Broadcast / Reduce. Effectively, pipelining hides the inta-node communications with the tree implementation, and therefore converges to the empirical bound with a pipeline with only $k = 4$ stages. On the other hand, pipelining provides algorithmic speedup with the ring implementation, and it requires up to $k = 32$ levels in the pipeline to saturate the throughput.  We observe similar behaviour across all of our test systems.

%A deeper pipeline is beneficial only for large (MB scale and up) message sizes. In theory, a deeper pipeline yields higher throughput because communications across all levels are overlapped. In practice, each stage in the pipeline poses a latency overhead, making deeper pipelines advantageous primarily for large message sizes.

%In (c) Broadcast, the pipelining matches NCCL’s performance at each message size with a different pipeline depth. NCCL uses an adaptive pipelining strategy based on a critical message length (e.g., 1 MB). As discussed in Section~\ref{sec:machine}, \codename{} offers an API for choosing the pipeline depth rather than having an adaptive optimization mechanism. %We left the choosing the optimal parameter to the user.

As a reference, we take MPICH (light blue) and NCCL (gold) library implementations as well as the aforementioned empirical bounds. Since NCCL does not offer Gather and Scatter collectives, we implement them directly with NCCL's point-to-point functions, as represented with the red curves. In (d) Reduce, even a deep pipeline falls short of NCCL's throughput. Remember that Reduce involves an additional computation that requires a GPU kernel call. NCCL successfully hides the computational overhead with a CUDA streaming mechanism. We do not apply CUDA-specific optimizations for the sake of generality.

\subsection{Scaling}
\begin{figure}[t]
\includegraphics[width=\columnwidth]{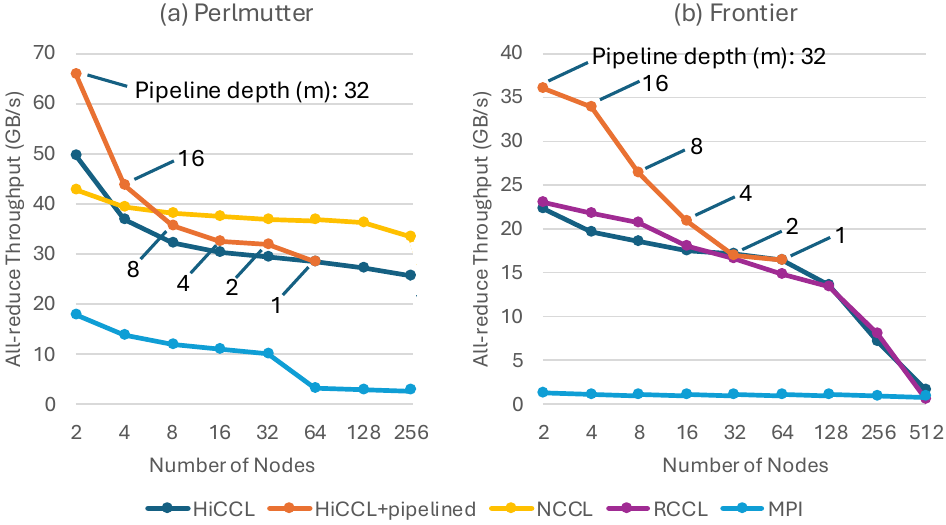}
\centering 
\caption{Scaling on (a) Perlmutter and (b) Frontier. HiCCL applies throughput-oriented optimizations, e.g., pipelining, that are effective up to 256 nodes. The performance difference between NCCL and HiCCL are mainly due to the kernel call overhead involved in reductions, where NCCL is more optimized.}
\label{fig:scaling}
\Description[<short description>]{<long description>}
\end{figure}

Figure~\ref{fig:scaling} demonstrates scaling on (a) Perlmutter and (b) Frontier. We employ All-reduce collective with two-step formulation (Section~\ref{sec:composite}) on both machines: We compose the collective only once with the proposed API and change the virtual hierarchy across machines. HiCCL is compared with MPI, NCCL, and RCCL, where available. For saturating the network in these experiments, buffer sizes were selected to be large (8.6 GB on Perlmutter and 17.2 GB on Frontier), which are determined based on the device memory capacity. Due to MPI's limitations with large buffer sizes~\cite{hammond2014int_max}, a 1 GB buffer size was utilized for MPI in the experiments on both machines.

The scaling experiments on Frontier reveal the the limitation of throughput-oriented optimizations we applied in this work. When scaled to more than 256 nodes (2,096 GPUs), the ratio of communication to computation drops significantly so that latency becomes the main bottleneck. In principle, latency-oriented collective design can be achieved with HiCCL's API, however, it is not in the scope of this work. Additionally, MPI on Frontier is tuned to minimize latency at large scale, and therefore preferable to any other communication library on more than 256 nodes.

\section{Related Work}

To the best of our knowledge, \codename{} is the first  library that offers optmized collective operations and performance portability across systems with Nvidia, AMD, and Intel GPUs.

%\subsection{Multi-NIC Striping}

Previous work ~\cite{coll2003using} implemented a multi-NIC striping algorithm with a lower-level communication API and for CPUs. For \codename{}, we applied the multi-NIC striping idea in the context of modern node architectures, where the endpoints are GPUs. We integrated the proposed striping algorithm with a high-level (GPU-aware) communication library, exploiting the fixed logical associations between GPUs and NICs as explained in Section~\ref{sec:multi_NIC}.

%\subsection{Hierarchical Communications}

%Hierarchical communications was previously applied~\cite{hidayetouglu2020petascale} specifically for Summit's multi-GPU node architecture. In that case, communication data is sparse and the sparsity pattern is specific to the imaging problem they are solving. In contrast, our work focuses on traditional collective functions with dense data, and expanding optimizations to multi-NIC nodes.

There is a body of previous work towards hierarchical collective communications. Nevertheless, they are either collective-specific~\cite{bienz2021modeling, lockhart2023characterizing}, system-specific~\cite{hidayetouglu2020petascale}, hardware-specific~\cite{graham2020scalable, graham2016scalable}, single-node~\cite{cai2021synthesizing}, for PCIe~\cite{kim2024tccl}, or CPU-only~\cite{luo2020han, peng2023optimizing}. \codename{} optimizes all collective functions with unified optimizations, achieves portability across architectures, and also demonstrates high throughput across systems of different GPU vendors (Nvidia, AMD, Intel). %Moreover, previous work relies on collective or point-to-point functions of a single library, whereas HiCCL is based on point-to-point functions of multiple GPU communication libraries for tailored implementation on a given system.

%\subsection{Composable Design}

A similar line of research (MSCCL~\cite{cai2021synthesizing, jangda2022breaking, shah2023taccl}) takes a different path on decomposition based generalized optimization by building a larger programming system with a domain-specific language and code generation. However, the previous work's scalability is limited up to 16 GPUs due to the costly code synthesis based on an SMT solver~\cite{de2008z3}. In contrast, \codename{} is a standalone runtime library that strives to be minimal and is scalable up to hundreds of nodes. Specifically, the initialization cost of HiCCL does not take more than six seconds on a thousand GPUs. %MSCCL's evaluation so far is limited to a few nodes.

%Moreover, \codename{} generalizes hierarchical communications to multiple---any---levels and pipeline depths to take advantage of exascale computer architecture (such as Frontier and Aurora) with multilevel hierarchies within and across nodes.

\section{Conclusion}

HiCCL is a collective communication library designed to exploit the hierarchical nature of modern network architectures.
HiCCL offers a compositional API with three primitives for building collective patterns and built-in transformations (striping, pipelining, and trading off between tree and ring communication topologies) for optimizing them. HiCCL relies on point-to-point functions of available libraries and IPC capabilities of vendors.

%HiCCL applies hierarchical factorization on the original pattern, and optimizes it for a described machine (Section~\ref{sec:optimization}).
%The generalization of hierarchical optimizations across collective communications and machines is HiCCL's highlight contribution.

We implemented HiCCL to be portable across Nvidia, AMD, and Intel GPUs and across various communication fabrics.
HiCCL improves the throughput of eight standard collective functions on four distinct machines with different architectures and vendors.
The geometric speedup of HiCCL is 17.0$\times$ over the native GPU-aware implementations on all systems, 1.15$\times$ over NCCL, 1.55$\times$ RCCL, and 12.1$\times$ OneCCL where available.
Our further evaluation demonstrates scalability up to 256 nodes (1,024--2,048 GPUs). 

\section{Acknowledgments}
This research was supported by the Exascale Computing Project (17-SC-20-SC), a collaborative effort of the U.S. Department of Energy Office of Science and the National Nuclear Security Administration.
This work was done on a pre-production supercomputer with early versions of the Aurora software development kit. This research used resources of the Argonne Leadership Computing Facility, a U.S. Department of Energy (DOE) Office of Science user facility at Argonne National Laboratory and is based on research supported by the U.S. DOE Office of Science Advanced Scientific Computing Research Program, under Contract No. DE-AC02-06CH11357.
This research used the Delta advanced computing and data resource which is supported by the National Science Foundation (award OAC 2005572) and the State of Illinois. Delta is a joint effort of the University of Illinois Urbana-Champaign and its National Center for Supercomputing Applications.
%This work is partially supported by Laboratory Directed Research and Development (Project Number: 2023-0104) funding from Argonne National Laboratory.
Sandia National Laboratories is a multimission laboratory managed and operated by National Technology \& Engineering Solutions of Sandia, LLC, a wholly owned subsidiary of Honeywell International Inc., for the U.S. Department of Energy’s National Nuclear Security Administration under contract DE-NA0003525.
This research used resources of the Oak Ridge Leadership Computing Facility at the Oak Ridge National Laboratory, which is supported by the Office of Science of the U.S. Department of Energy under Contract No. DE-AC05-00OR22725.
This research used resources of the National Energy Research Scientific Computing Center (NERSC), a Department of Energy Office of Science User Facility using NERSC award ASCR-ERCAP0029675.

\bibliographystyle{plain}
\bibliography{references}

\end{document}